\documentclass{article}

\usepackage{arxiv}

\usepackage[utf8]{inputenc} 
\usepackage[T1]{fontenc}    
\usepackage{hyperref}       
\usepackage{url}            
\usepackage{booktabs}       
\usepackage{amsfonts}       
\usepackage{nicefrac}       
\usepackage{microtype}      
\usepackage{lipsum}
\usepackage{graphicx}
\graphicspath{ {./images/} }

\title{Design of a validation methodology for a prototype wristband for capturing muscle signals and upper limb movement}

\author{
  Miguel Alejandro Mantilla\\
  Universidad de Antioquia UdeA\\
  Medellín, Colombia, 050010 \\
  \texttt{ana.montanez@udea.edu.co} \\
   \And
  Ana Maria Montañez \\
  Universidad de Antioquia UdeA\\
  Medellín, Colombia, 050010 \\
  \texttt{miguel.mantilla@udea.edu.co} \\
  \And
Daniel Escobar Saltarén\\
  Universidad de Antioquia UdeA\\
  Medellín, Colombia, 050010 \\
  \texttt{daniel.escobars@udea.edu.co} \\
  \And
Sofía C. Henao\\
  Tecnologico de Monterrey\\
  School of Engineering and Sciences\\
  64849, Monterrey, N.L, Mexico \\
  \texttt{sofia.henao@tec.mx} \\
  \And
  María B. Salazar-Sánchez\\
  Universidad de Antioquia UdeA\\
  Medellín, Colombia, 050010 \\
  \texttt{bernarda.salazar@udea.edu.co} \\
}

\begin{document}
\maketitle
\begin{abstract}
Surface electromyography (sEMG) is a non-invasive technique that is widely used to control myoelectric prostheses and other human–machine interfaces. However, the high cost of commercial systems can limit accessibility in academic, research, and industrial settings, particularly in developing countries. To address this issue, the development of low-cost acquisition systems has been promoted, which requires assurance of electrical safety and the reliability of the acquired signals. Against this backdrop, this article presents a validation protocol based on international standards, including IEC 60601 and ANSI/AAMI EC13. Based on these standards, electrical safety tests are implemented, including measuring leakage currents under normal conditions, evaluating double insulation, and verifying continuity between electrodes and circuits. Additionally, the article incorporates functional evaluations of the system by comparing signals acquired from an eight-electrode sEMG wristband prototype with those from a commercial reference device (PortiLab2). This comparison uses statistical metrics such as Pearson's correlation, Bland–Altman analysis, and mean squared error. These evaluations are expanded to include tests of signal stability under resting and contraction conditions, validation of UART and Bluetooth communication, frequency response analysis, mechanical characterization of the casing, and evaluation of user comfort. The results show leakage currents ranging from 11.4 µA to 13.5 µA, slightly above the established threshold, and a bias current close to the regulatory limit. Adequate insulation, clear differentiation between muscle signals and noise, and high correlation with the reference system (r > 0.85) were demonstrated, along with stable, loss-free data transmission. One limitation is that constraints were identified at the power-supply stage, necessitating the use of an external power source during wireless testing. In addition, discrepancies were observed in the frequency response relative to the simulation, particularly at high-gain stages. The casing exhibited elastic behavior under loads up to 98 N. These results enable us to establish a minimum technical and functional validation framework for low-cost sEMG systems intended for research and training. Furthermore, the proposed protocol can be implemented using basic laboratory equipment by a single operator, providing a practical, reproducible, and accessible methodology for validating biomedical devices in research settings. 
\end{abstract}


\section{Introduction}
Surface electromyography is a non-invasive technique used to record the electrical activity generated by muscles during contraction, thereby enabling the identification of users' movement intentions for controlling myoelectric prostheses and other human–machine interfaces \cite{Madhavan2005-xr} \cite{Liu2024-vt}. This signal is widely applied in areas such as rehabilitation, neuromuscular diagnosis, and biomechanical analysis, due to its ability to capture neural information in real time without requiring invasive procedures \cite{Sensinger2019-ko}. Consequently, the acquisition of sEMG signals, often combined with motion data, has gained increasing relevance in the development of portable and wearable systems.

However, access to commercial equipment for the reliable acquisition of sEMG signals remains limited, particularly in academic environments or settings with limited budgets. Devices such as Delsys Trigno and Cometa® systems can exceed USD 10,000 per unit \cite{Bawa2022-lz}, restricting their implementation to well-resourced institutions. Additionally, costs associated with accessing international standards required for device validation, such as IEC 60601 and ANSI/AAMI EC13 \cite{Acevedo_Lopez2015-wa}, further limit their adoption. As a result, numerous low-cost sEMG acquisition prototypes, particularly wearable armbands, have been developed in academic and research settings.
These prototypes are generally not intended for clinical use and lack formal certification. Nevertheless, given their direct contact with human skin, ensuring electrical safety and the reliability of acquired signals is essential. This lack of standardization often results in uncertainty about their technical performance and safe operation. Furthermore, previous studies have shown that a significant proportion of adverse events related to medical devices are associated with technical failures rather than clinical use, highlighting the need for systematic and objective evaluation frameworks \cite{Correa2017-tj}.

In addition, in countries such as Colombia, the limited availability of certified laboratories capable of conducting this type of validation represents an additional barrier to the development and safe implementation of emerging biomedical technologies \cite{Acevedo_Lopez2015-wa}. In this context, the development of structured and accessible validation protocols is crucial not only to support regulatory alignment but also to ensure reliable device performance under real-world conditions.

This article presents the technical characterization and validation of a custom-designed modular 8-channel sEMG armband equipped with dry electrodes and an Inertial Measurement Unit (IMU). The core contribution of this work is the design and implementation of a replicable validation protocol based on international standards such as IEC 60601-1, which establishes safety requirements for medical electrical equipment. The proposed protocol evaluates electrical safety through leakage current measurements, signal quality compared to a commercial reference device (PortiLab2), structural integrity through mechanical testing, and data transmission stability. In addition, it incorporates assessments of signal stability, synchronization, data integrity, ergonomics, and usability. The objective is to establish a clear methodological framework for evaluating similar low-cost sEMG systems and to facilitate their implementation in applied research and training contexts in bioengineering \cite{Farina2014-kl}.

\section{Materials and Methods}
\label{sec:headings}
\subsection{Device Description}
The sEMG armband features a modular structure to facilitate manufacturing, evaluation, and adaptability to different usage contexts. It incorporates eight sEMG electrodes integrated into an elastic silicone band, allowing adjustment for different forearm sizes. Each sensing unit consists of metallic electrodes arranged in a differential configuration and housed within 3D-printing enclosures to protect the internal circuitry and ensure stable mechanical attachment.
Additionally, the device uses an ESP32-S3 \cite{espressif_esp32s3_2026} microcontroller to handle data acquisition, processing, and transmission. The analog-to-digital conversion of the sEMG signals is performed using an ADS1015 \cite{ti_ads101x_2025} converter, which provides a 14-bit resolution and a sampling rate of up to 800Hz per channel. Furthermore, the system incorporates an IMU \cite{tdk_icm42605} that provides complementary information about acceleration and forearm orientation. This information is useful in applications where a correlation between movement and muscular activity is required.

The EMG signal is conditioned through a pre-filtering stage, followed by amplification with an AD620 instrumentation amplifier \cite{Devices2003-te}. Subsequently, band-pass filtering is applied, consisting of second-order active Butterworth high-pass and low-pass filters, along with a second-order twin-T notch filter to attenuate power-line interference. Additionally, an analog processing stage is implemented to generate a rectified envelope signal, which is useful for representing muscle contraction intensity in real time.

The system supports data transmission through two interfaces: a Universal Asynchronous Receiver–Transmitter (UART) for wired communication and Bluetooth Low Energy (BLE) for wireless transmission. A graphical user interface developed in Python enables real-time signal visualization, data storage in .xlsx format, and basic system configuration, facilitating its use in validation environments and pilot testing.

\subsection{Normative Reference}
Armband validation was conducted in accordance with international standards that establish safety and performance requirements for medical electrical devices. In particular, IEC 60601-1 was used as a reference standard because it defines essential safety requirements for electromedical equipment, including protection against electrical hazards, limits on leakage currents, insulation requirements, classification of applied parts, and operation under normal conditions. Additionally, IEC 60601-2-40 focuses on stimulation devices and neuromuscular monitoring; therefore, aspects related to signal stability and frequency response across different frequency ranges were considered in the validation protocol.

IEC 61010-1, which addresses safety requirements for measurement and laboratory equipment, was considered during the design of tests involving signal generators and oscilloscopes. Finally, ANSI/AAMI EC13 establishes verification procedures for biomedical monitoring devices, focusing on evaluating of accuracy, stability, and performance under controlled conditions. The proposed protocol is not intended to replace formal certification; however, it provides a minimum framework for technical and functional validation that can be applied in academic, research, and development environments.

\subsection{Validation Protocol}
\begin{figure}[h]
  \centering
  \includegraphics[width=1\textwidth]{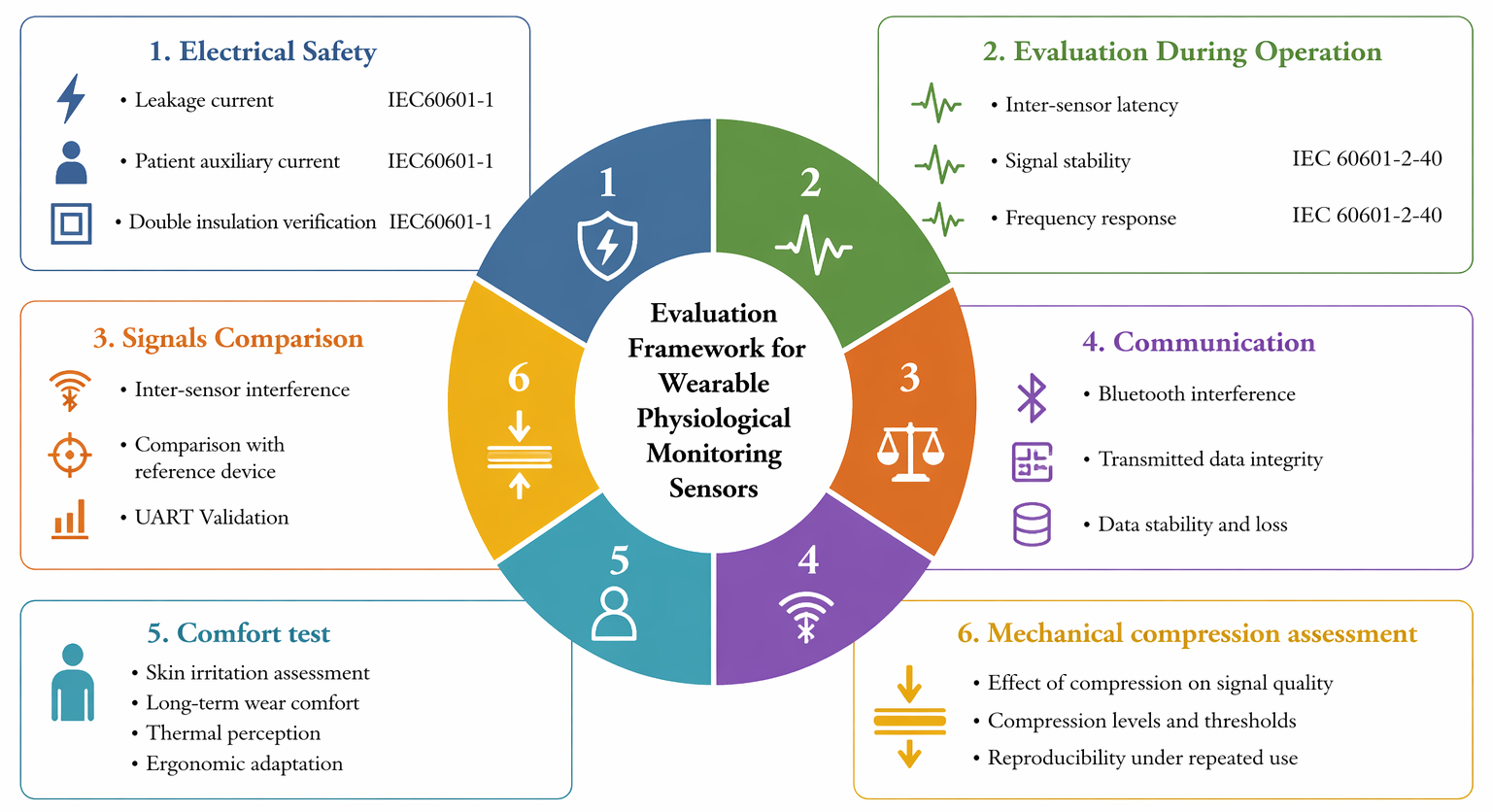}
  \caption{Overview of the proposed validation protocol for the sEMG armband.}
  \label{fig:fig1}
\end{figure}

As illustrated in Figure \ref{fig:fig1}, the proposed validation framework was structured into six main evaluation categories to comprehensively assess the performance, safety, and usability of the developed system. These categories include: (1) electrical safety evaluation based on IEC 60601-1 requirements; (2) operational performance assessment, including latency, stability, and frequency response according to IEC 60601-2-40; (3) signal comparison with reference acquisition systems to verify signal quality and reliability; (4) communication assessment to evaluate data transmission performance; (5) comfort evaluation focused on user experience during device use; and (6) mechanical compression evaluation to analyze the physical interaction between the device and the user.

\subsubsection{Electrical safety}
Electrical safety evaluation is a critical stage in the validation of biomedical acquisition systems, as it verifies that the device operates within safe limits for human use under normal operating conditions. In this study, leakage current measurements were performed to quantify the current that could flow through the human body during device operation, particularly in systems involving direct or indirect electrical contact with users.

To conduct the evaluation, an experimental circuit was implemented to emulate the electrical impedance of the human body by connecting a 1~k$\Omega$ resistor between the signal line and ground, following standard practices in medical electrical safety testing. The voltage drop across the resistor was measured, and the leakage current was subsequently calculated using Ohm’s law.

To ensure measurement reliability and minimize random variability, each sensor configuration was evaluated at least three times independently, in accordance with the recommendations of ISO 14971 for medical device risk management and verification processes. The resulting measurements were analyzed to verify compliance with acceptable electrical safety limits.

Additionally, the patient auxiliary current, defined as the current required for proper electrode operation, was evaluated using the same experimental setup to ensure compliance with established safety thresholds. A double-insulation test was also performed to verify adequate electrical isolation between the device’s internal circuitry and the user. The procedures described in this section were based on the requirements established in IEC 60601-1.

\subsubsection{Evaluation during operation}

The operational evaluation was designed to assess the performance and reliability of the acquisition system under controlled operating conditions. This stage focused on analyzing signal stability, internal noise behavior, and frequency response characteristics to verify that the system maintained adequate performance during continuous operation. These evaluations are particularly important in biomedical acquisition systems, where signal integrity and stability directly influence the accuracy and reliability of physiological measurements.

As an initial step, a signal stability assessment was performed by recording the system output in the absence of a biological load. This procedure allowed the identification of internal interference sources, baseline fluctuations, and electronic noise generated by the acquisition circuitry itself, without the influence of physiological signals. The recorded data were analyzed using descriptive statistics, including the mean, standard deviation, and coefficient of variation, to quantify signal variability and evaluate overall stability. To improve the robustness and reproducibility of the analysis, the evaluation was conducted using at least three different electrodes, providing sufficient data for comparative assessment across multiple configurations.

In addition to stability analysis, the system's frequency response was evaluated using a waveform generator to apply controlled input signals across different frequency ranges. This procedure enabled the characterization of the electrode and acquisition system behavior under varying signal conditions commonly encountered in biomedical applications. The generated signals were applied to the system input, and the corresponding output responses were recorded and analyzed to determine signal attenuation, distortion, and consistency across frequencies. Experimental results were subsequently compared with theoretical and simulation-based responses to verify the accuracy and expected performance of the proposed design.

The procedures implemented in this section were primarily based on the recommendations established in IEC 60601-2-40 for medical electrical equipment related to electromyographic and physiological signal acquisition systems. However, certain methodological adaptations were introduced to accommodate the characteristics of the developed prototype and the study's experimental conditions.

\subsubsection{Signal comparison}
The signal comparison stage was designed to evaluate the temporal synchronization, interference behavior, and overall signal fidelity of the proposed acquisition system in relation to a commercial reference device. These analyses are essential to ensure that the recorded signals preserve their temporal consistency and maintain comparable quality across all acquisition channels, particularly in applications involving multichannel biomedical monitoring.

To verify channel synchronization and characterize potential acquisition delays, latency analyses were performed across electrodes. For this purpose, a square-wave stimulus was simultaneously applied to each sensor, allowing the identification of temporal phase shifts or lag between channels. Since the acquisition system uses sequential sampling, selected sensor pairs were analyzed to determine whether the sampling architecture introduced measurable timing differences during signal acquisition. This evaluation enabled assessment of the system's temporal alignment accuracy and its suitability for synchronized multichannel recordings.

Additionally, interference behavior between channels was evaluated to determine the degree of signal coupling or cross-interference within the acquisition circuitry. During this procedure, a controlled stimulus was applied independently to each electrode while the responses of the remaining sensors were monitored. The resulting measurements enabled assessment of whether activating one channel caused unwanted effects or distortions in neighboring channels, thereby providing information on channel isolation and system robustness.

To further validate signal quality and acquisition fidelity, simultaneous recordings were obtained with both the developed prototype and a commercial reference device under identical experimental conditions. The acquired raw signals were subsequently processed to extract representative signal features commonly employed in electromyographic analysis, including root-mean-square (RMS), mean absolute value (MAV), variance (VAR), and waveform length (WL). These features were selected because they provide quantitative information related to signal amplitude, variability, and morphological behavior.

The extracted features from both acquisition systems were compared on a window-by-window basis using quantitative performance metrics. Specifically, the mean absolute percentage error (MAPE) was used to quantify the relative difference between systems, while Pearson’s correlation coefficient was used to assess the degree of linear similarity between the corresponding signals. Together, these analyses enabled the assessment of signal stability, fidelity, and agreement between the proposed system and the commercial reference device. The validation procedures described in this section were primarily informed by technical and experimental experience with biomedical signal acquisition systems.
\subsubsection{Communication}
The communication evaluation was designed to assess the reliability and stability of both wired and wireless data transmission implemented in the proposed acquisition system. Specifically, the analysis focused on the performance of UART-based wired communication and Bluetooth Low Energy (BLE) wireless transmission, which are fundamental for ensuring accurate real-time acquisition and monitoring of biomedical signals.

The primary objective of this evaluation was to verify that the transmitted data were received continuously and in real time without corrupted characters, packet losses, synchronization failures, or structural inconsistencies in the communication frames. Maintaining data integrity during transmission is particularly important in biomedical applications, where transmission errors may compromise signal interpretation and subsequent analysis.

For the wired communication assessment, UART transmission stability was analyzed by continuously monitoring the received data stream under normal operating conditions. The transmitted signals were inspected to identify possible communication interruptions, incomplete frames, or encoding errors that could affect signal reconstruction and processing.

In the case of wireless communication, additional tests were conducted to assess the robustness of BLE transmission in environments with potential electromagnetic or wireless interference. To simulate realistic operating conditions, multiple Bluetooth-enabled devices were activated simultaneously during continuous data transmission sessions lasting approximately one minute. During these tests, the system was monitored to determine whether external wireless activity introduced delays, packet losses, transmission instability, or signal corruption.

The overall communication performance was subsequently analyzed based on transmission continuity, data integrity, and the system’s ability to maintain stable real-time operation under both isolated and interference-prone conditions. The methodology described in this section was developed based on experimental experience and previous studies reported by \cite{Acevedo_Lopez2015-wa}, \cite{Van_Lier2020-vk}, and \cite{Fuentes_Del_Toro2019-yh}.

\subsubsection{Comfort test}
User comfort was evaluated as part of the usability and ergonomic assessment of the wearable device, following the recommendations established in IEC 62366-1 for usability engineering in medical devices, together with general ergonomic principles commonly applied to wearable systems. This evaluation aimed to determine whether the device could be worn for prolonged periods without causing discomfort, excessive pressure, movement restriction, or adverse skin effects during typical use conditions.

The assessment was conducted under two different scenarios: (i) a resting condition and (ii) an active condition involving activities of daily living associated with forearm movement. In both cases, the device was positioned directly on the user’s forearm and worn continuously throughout the evaluation period to analyze its physical interaction with the user under static and dynamic conditions.

During the resting condition, the participant remained seated for approximately 15 minutes while wearing the device in a relaxed posture. Throughout this period, observations focused on perceived comfort, overall fit, localized pressure points, skin irritation, and any discomfort from prolonged contact between the device and the forearm. Particular attention was given to identifying whether the mechanical structure or fastening mechanism generated excessive compression or uneven pressure distribution.

For the active condition, the participant performed repetitive forearm movements for approximately 10 minutes, including flexion, extension, and rotational motions representative of common daily activities. This stage of the evaluation was intended to analyze the device's behavior during movement and to determine its ability to maintain adequate positioning without compromising user mobility. Factors such as device stability, slippage, skin friction, restricted movement, and the need for repositioning or readjustment were carefully assessed throughout the procedure.

Following each test, a qualitative visual inspection of the skin was conducted to identify possible marks, redness, or signs of excessive pressure and friction caused by the wearable structure. In addition, descriptive user feedback was collected to document subjective perceptions related to comfort, usability, and overall wearing experience. The collected observations were subsequently used to identify potential ergonomic improvements and guide future iterations of the device design.

\subsubsection{Mechanical compression assessment}
A mechanical compression assessment was performed to evaluate the structural integrity and mechanical resistance of the sensor enclosure under loading conditions representative of normal use. This evaluation is relevant for wearable biomedical devices, as the enclosure must withstand external forces generated during handling, placement, adjustment, and daily operation without compromising user safety or device functionality. The procedure was conducted in accordance with the general mechanical safety considerations described in IEC 60601-1, which establishes that device enclosures must maintain their structural integrity under expected mechanical stresses.

The compression test was carried out using a universal testing machine (Shimadzu AG-X Plus), which allowed controlled application and monitoring of mechanical loads. During the experiment, a progressively increasing vertical compressive force was applied directly to the enclosure until reaching a maximum load of 98~N, equivalent to approximately 10~kg. This value was selected as an estimated upper limit of the contact and compression forces that could reasonably occur during typical device use and handling conditions.

Throughout the test, force and displacement were continuously recorded to characterize the structure's mechanical response under compression. Using these measurements, a stress--strain curve was then constructed to analyze the enclosure’s deformation behavior. Stress values were calculated using the enclosure's cross-sectional area, and strain was determined as the ratio of the measured displacement to the structure's original height.

The analysis focused on identifying elastic and plastic deformation, as well as any evidence of permanent structural damage or failure after load removal. Particular attention was given to determining whether the enclosure could recover its original geometry after compression and whether the applied loads produced fractures, excessive deformation, or instability in the mechanical structure. Finally, the experimental results were compared with the typical mechanical properties reported for PETG to estimate the safety margin and evaluate the suitability of the enclosure design for wearable biomedical applications.

\section{Results}
This section presents the experimental results obtained by applying the proposed validation protocol to the developed eight-electrode sEMG armband. The analyses were organized according to the different evaluation stages defined in the protocol, including electrical safety, operational performance, signal comparison, communication reliability, user comfort, and mechanical compression assessment. Together, these results provide a comprehensive evaluation of the device with respect to safety, signal quality, structural integrity, and usability under representative operating conditions.

\subsection{Electrical safety}
To evaluate the leakage current, ten independent repetitions were performed for each electrode configuration, and the obtained results are summarized in Table \ref{tab:table1}. According to IEC 60601-1, the maximum permissible leakage current under normal operating conditions is 10~µA, and although the measured mean values slightly exceeded this limit, the deviation remained minimal and close to the acceptable threshold. These differences may be related to experimental conditions and limitations associated with the measurement setup, including environmental noise, grounding conditions, and minor variations in the acquisition circuitry during testing. Overall, the results indicate near-compliant electrical safety performance for the proposed system and suggest that further optimization of the electronic design could reduce leakage current levels.

\begin{table}[h]
 \caption{Leakage Current Measurements Across Repetitions ($\mu A$).}
  \centering
  \begin{tabular}{llllll}
    \toprule
    \multicolumn{6}{c}{Leakage current ($\mu$A)}               \\
    \cmidrule(r){2-6}
    Sensor & 1 & 2 & 3 & 4 & Mean ± std\\
    \midrule
    1 & 15,36	& 15,36	& 16,98	& 20,62 &	17,08 ± 2,15  \\
    2 &	15,77 &	15,77 &	19,98 &	20,39 &	17,98 ± 2,21 \\
    3 &	15,77 &	15,77 &	16,98 &	18,76 &	16,82 ± 1,22 \\
    4 &	15,04 &	15,04 &	26,68 &	17,95 &	18,68 ± 4,77 \\
    5 &	16,01 &	16,01 &	24,66 &	17,95 &	18,66 ± 3,56 \\
    6 &	17,71 &	17,71 &	20,62 &	17,95 &	18,50 ± 1,23 \\
    7 &	17,63 & 21,83 &	20,38 &	20,62 &	20,12 ± 1,54 \\
    8 &	21,83 &	17,63 &	24,42 &	17,95 &	20,46 ± 2,83\\
    \bottomrule
  \end{tabular}
  \label{tab:table1}
\end{table}

In addition to leakage current analysis, the patient auxiliary current was also evaluated, since this parameter is directly related to the current required for proper electrode operation during signal acquisition. IEC 60601-1 establishes a maximum allowable limit of 100~µA for this parameter, and, as shown in Table \ref{tab:table2}, the measured values exhibited moderate variability among the evaluated electrodes, with an overall mean of 101.03 $\pm$ 41.52~µA. Although the average value remained close to the regulatory limit, several individual measurements exceeded the recommended threshold, suggesting variability in the current distribution across channels. These results indicate that further optimization of the electronic design and acquisition circuitry may be necessary to ensure full compliance with the established safety requirements.

\begin{table}[h]
 \caption{Patient auxiliary current ($\mu A$).}
  \centering
  \begin{tabular}{ll}
    \toprule
    \cmidrule(r){1-2}
    Repetition & Patient auxiliary current \\
    \midrule
    1 &	135,12\\
    2 &	135,12\\
    3 &	170,70\\
    4 &	152,18\\
    5 &	73,320\\
    6 &	63,470\\
    7 &	59,760\\
    8 &	59,660\\
    9 &	93,300\\
    10 &	67,690\\
    Mean &	101,030\\
    \bottomrule
  \end{tabular}
  \label{tab:table2}
\end{table}

\begin{figure}
  \centering
  \includegraphics[width=0.6\textwidth]{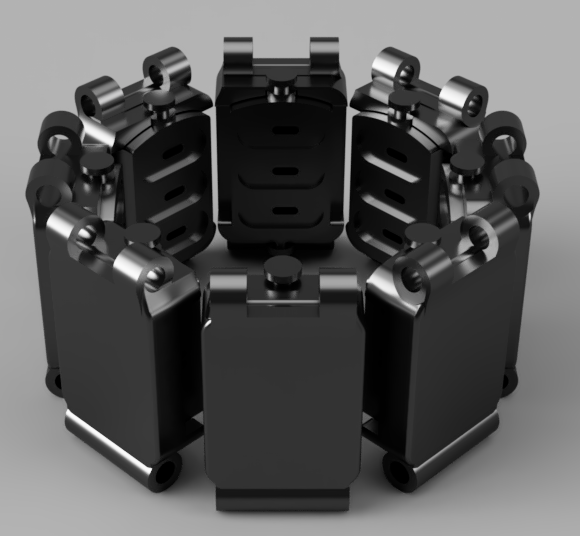}
  \caption{Modular armband prototype with integrated sensing units.}
  \label{fig:fig2}
\end{figure}

It was also essential to ensure that no electrical circuitry remained in direct contact with the patient’s skin during device operation. As shown in Figure \ref{fig:fig2}, the sEMG electrode circuits were enclosed within individual protective housings, while the battery and microcontroller were integrated into a separate enclosure, thereby minimizing both intentional and accidental access to the electronic components and improving overall user safety.

\subsection{Evaluation during operation}
The three repetitions presented in Table \ref{tab:table3} showed consistent behavior, with mean values close to 1~mV and low relative dispersion across measurements. These results indicate stable system performance and confirm that, with the bracelet removed from the arm and in the absence of an external load, the acquisition system does not introduce significant internal interference or noise.

\begin{table}[h]
 \caption{Descriptive statistics of signal measurements per repetition (mV). CV: coefficient of variation; SD: standard deviation.}
  \centering
  \begin{tabular}{lllll}
    \toprule
    \cmidrule(r){1-5}
    Repetition & Mean (mV) & SD(mV) & CV (\%) & Mean variation (\%)\\
    \midrule
    1 &	1,0003 &	0,0175 &	1,7500 &	0,0600\\
    2 &	1,0003 &	0,0248 &	2,4800 &	0,0900\\
    3 &	1,0008 &	0,0276 &	2,7600 &	0,1500\\
    Mean &	1,0005 &	0,0233 &	2,3300 &	0,1000\\
    \bottomrule
  \end{tabular}
  \label{tab:table3}
\end{table}

\begin{figure}[h]
  \centering
  \includegraphics[width=0.9\textwidth]{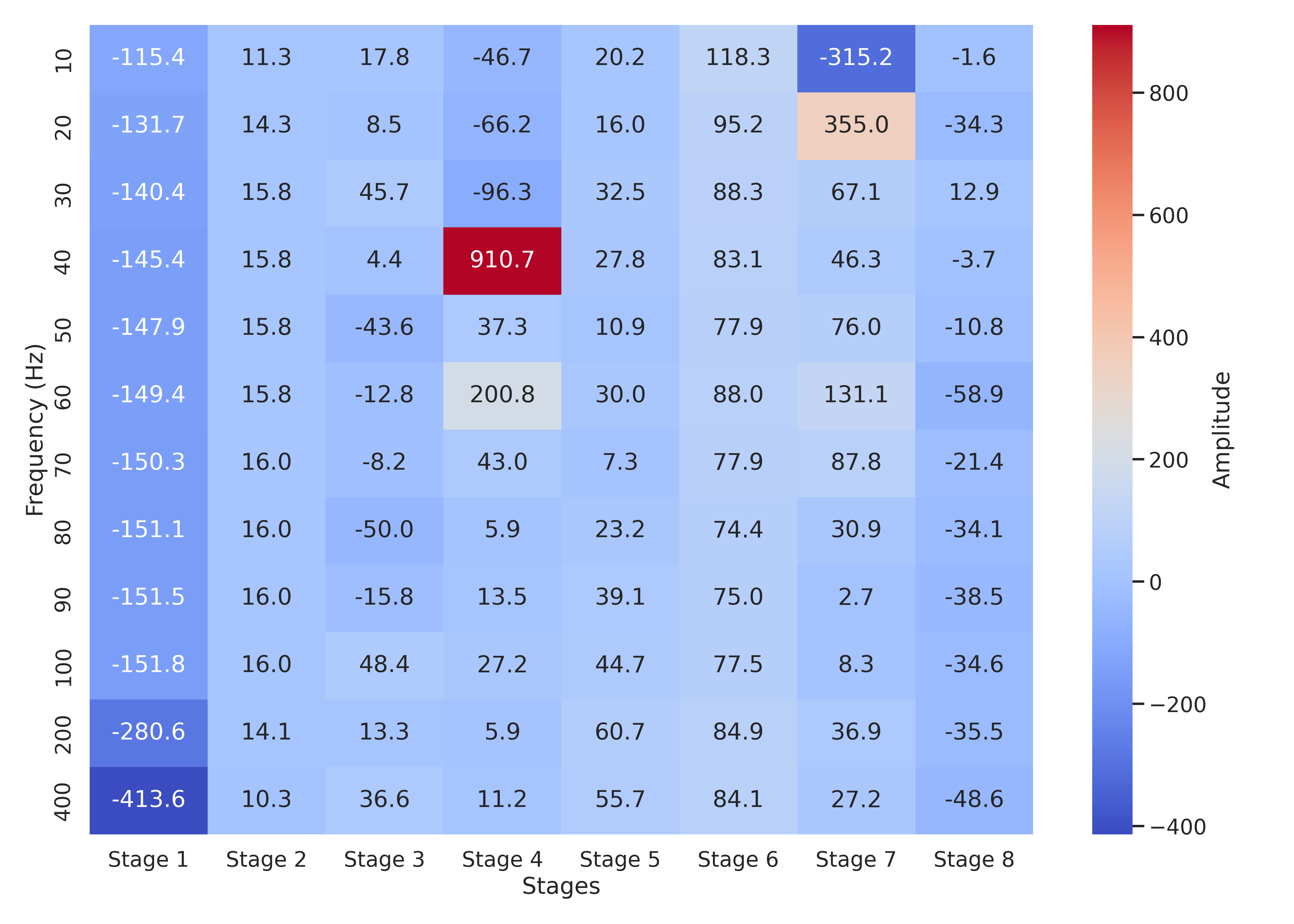}
  \caption{Heatmap-based visualization of percentage error across stages and frequencies (\%).}
  \label{fig:fig3}
\end{figure}

On the other hand, Figure \ref{fig:fig3} presents the percentage error obtained by comparing the simulated and experimental behavior of each stage of the sEMG sensor circuit. The circuit was divided into different processing stages, including the preamplification stage (Stage 1), instrumentation amplifier (Stage 2), notch filter (Stage 3), high-pass filter (Stage 4), band-pass filter (Stages 5--7), and rectifier stage (Stage 8), allowing for a more detailed evaluation of the performance of each component within the acquisition chain.

The results show that the preamplification stage (Stage 1) exhibited the highest percentage errors, whereas the instrumentation amplifier (Stage 2) demonstrated more stable behavior with lower deviations between simulated and experimental responses. Similarly, the notch filter (Stage 3) and band-pass filter (Stage 5--7) stages presented variations that depended on the analyzed frequency range, while the rectification stage (Stage 8) showed moderate discrepancies likely associated with the non-ideal behavior of practical electronic components.

In contrast, the high-pass filter stage (Stage 4) exhibited substantially larger errors than the other stages, with deviations as high as 911\%. This behavior may be associated with experimental issues during testing, such as unstable connections, grounding inconsistencies, component tolerances, or differences between the simulated and physical implementations of the circuit, particularly at low-frequency operation.

Despite these differences, the overall system demonstrated functional and stable performance, with experimental responses generally following the expected trends predicted by simulation. The observed deviations can be mainly attributed to experimental conditions and the inherent limitations of real electronic components. In particular, the presence of negative percentage errors indicates that the experimental measurements were lower than the simulated values, which may be explained by factors such as resistor tolerances, non-ideal operational amplifier characteristics, parasitic effects introduced during physical implementation, measurement losses, and environmental electrical noise affecting the acquisition setup.

\subsection{Signal comparison}
\begin{figure}[h]
  \centering
  \includegraphics[width=0.8\textwidth]{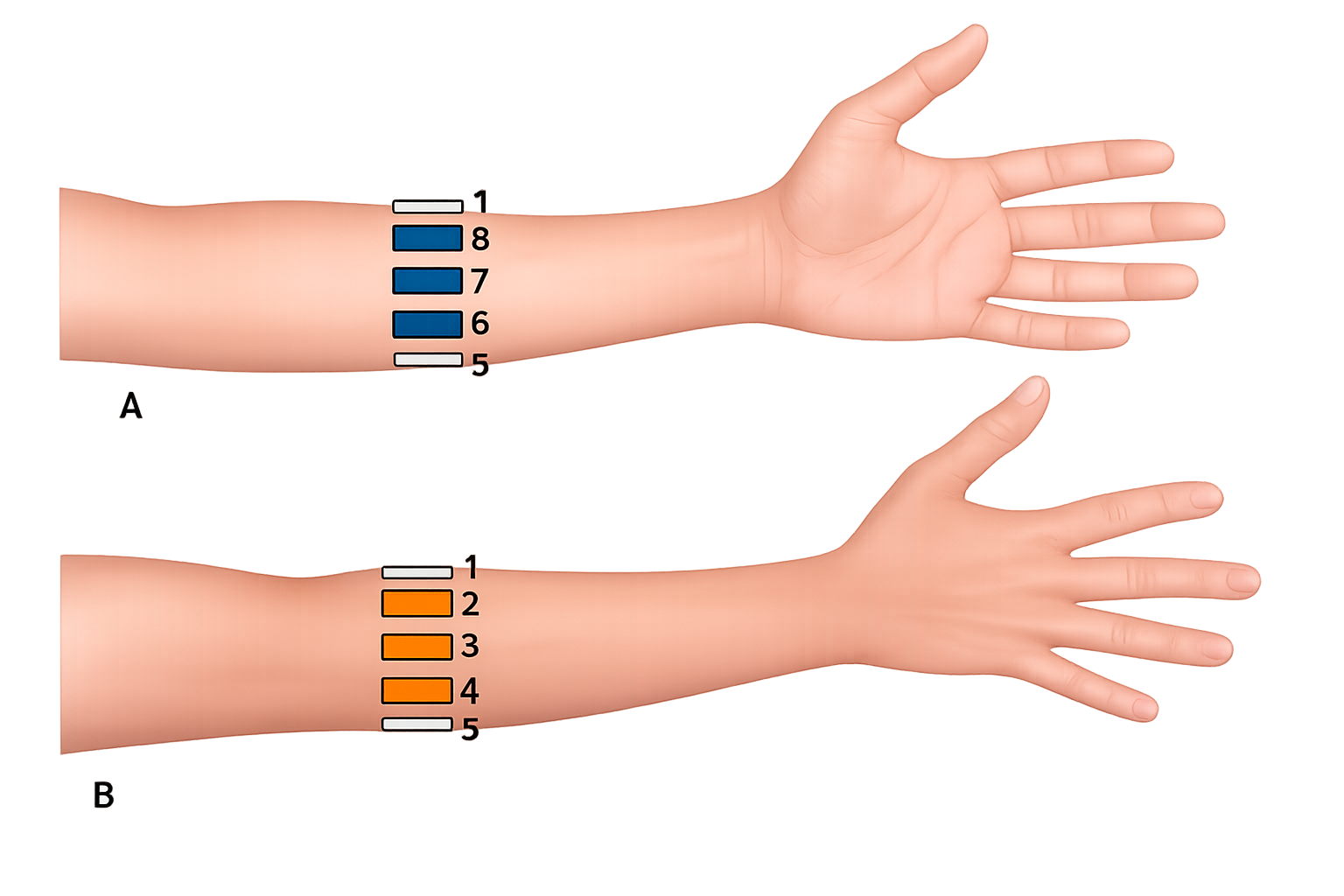}
  \caption{Spatial distribution of sEMG sensors.  (A) Anterior view (B) Posterior view.}
  \label{fig:fig4}
\end{figure}

Since the ADS1015 performs serial data acquisition rather than simultaneous channel sampling, temporal differences may occur between sensor readings during operation. To evaluate the potential impact of this sequential acquisition process, sensors 2, 4, and 8 were selected for analysis due to their spatial distribution across the armband, as illustrated in Figure \ref{fig:fig4}. Specifically, these electrodes were positioned at different locations around the forearm to provide a representative assessment of synchronization behavior and timing variations between channels acquired sequentially. 

The electrode arrangement included both anterior (A) and posterior (B) forearm configurations, enabling the evaluation of the multichannel acquisition system under different spatial distributions of the sEMG sensors. The selected sensor configurations were evaluated to determine the stability and consistency of the acquisition process under sequential sampling conditions. The descriptive statistics obtained from the three experimental repetitions are summarized in Table \ref{tab:table4}. Overall, the measurements were highly consistent across repetitions, with mean signal amplitudes remaining close to 1~mV in all cases, indicating stable signal acquisition regardless of the spatial location of the evaluated electrodes.

The variability observed between repetitions remained low throughout the experiments, the standard deviation values were below 0.03~mV for all measurements, reflecting limited fluctuations in the recorded signals. Likewise, the coefficient of variation (CV) ranged from 1.75\% to 2.76\%, suggesting low relative dispersion with respect to the mean amplitude values. The mean variation percentage also remained minimal, with an overall average of 0.10\%, further supporting the repeatability and consistency of the measurements obtained from the distributed electrode configurations.

\begin{table} [h]
 \caption{Descriptive statistics of signal measurements per repetition (mV). CV: coefficient of variation; SD: standard deviation.}
  \centering
  \begin{tabular}{lllll}
    \toprule
    \cmidrule(r){1-5}
    Repetition & Mean (mV) & SD(mV) & CV (\%) & Mean variation (\%)\\
    \midrule
    1 &	1,0003 &	0,0175 &	1,7500 &	0,0600\\
    2 &	1,0003 &	0,0248 &	2,4800 &	0,0900\\
    3 &	1,0008 &	0,0276 &	2,7600 &	0,1500\\
    Mean &	1,0005 &	0,0233 &	2,3300 &	0,1000\\
    \bottomrule
  \end{tabular}
  \label{tab:table4}
\end{table}

These results indicate that the proposed acquisition system maintained stable electrical performance despite the ADS1015's sequential sampling strategy and the spatial separation between electrodes. In general, low variability and consistent signal amplitudes suggest that the system introduces limited internal interference while preserving reliable multichannel acquisition performance under the evaluated conditions.

\begin{figure}[h]
  \centering
  \includegraphics[width=0.8\textwidth]{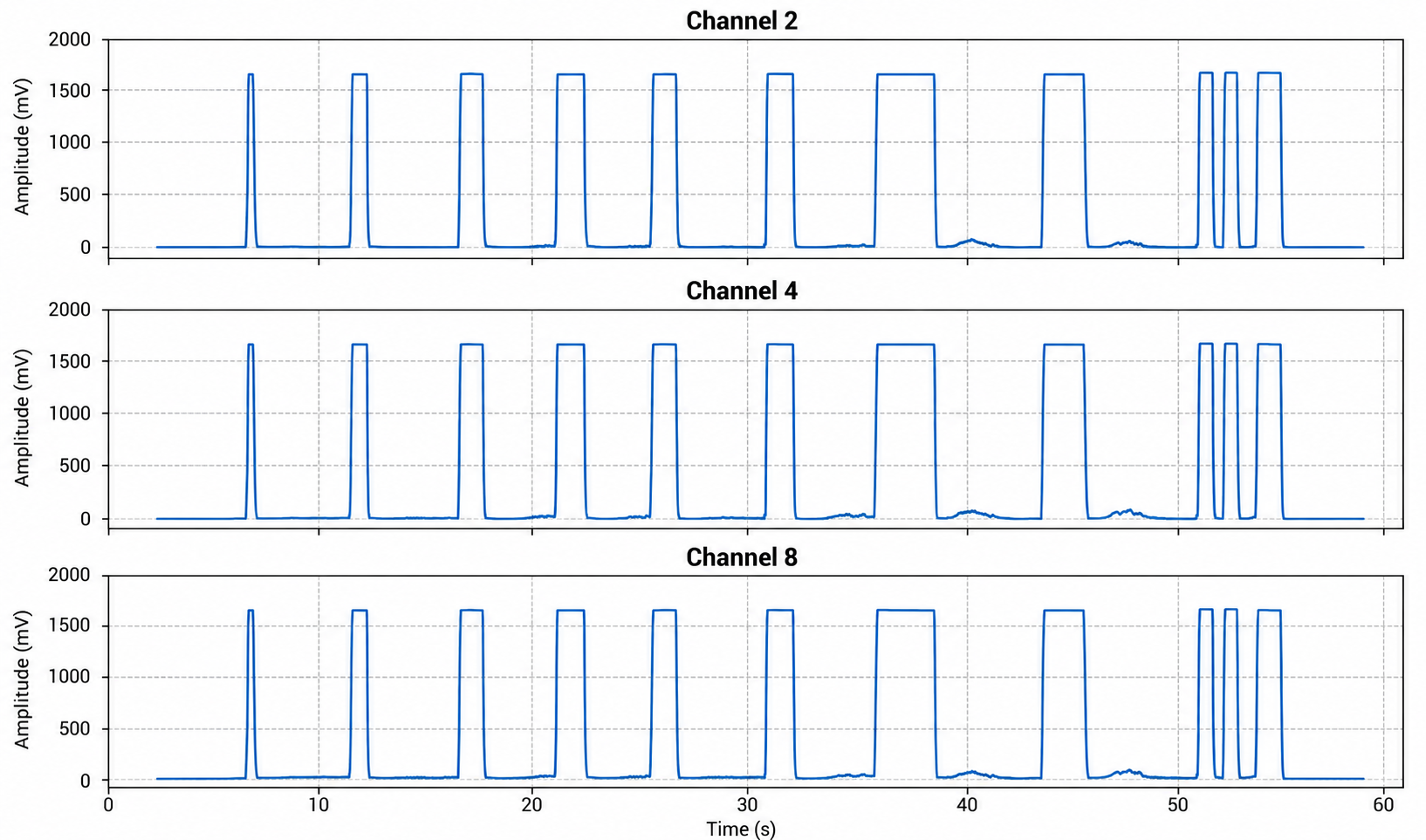}
  \caption{Signals recorded in channels 2, 4, and 8 after the simultaneous application of a stimulus during the latency test.}
  \label{fig:fig5}
\end{figure}

Figure \ref{fig:fig5} presents the recorded signals from channels 2, 4, and 8, in which a clear temporal alignment of the rising edges can be visually observed following stimulus application. This behavior suggests consistent synchronization among the evaluated channels despite the acquisition system's sequential sampling strategy. To complement the visual analysis, Table \ref{tab:table5} summarizes the detection times, expressed in milliseconds, at which each channel exceeded the predefined threshold, together with the corresponding temporal differences between consecutive channel pairs (Channel 2 vs. Channel 4 and Channel 4 vs. Channel 8).

Table \ref{tab:table5} showed that all channels detected the stimulus within a margin equivalent to a single sampling interval, indicating minimal temporal displacement between recordings. Consequently, no significant latency or synchronization errors were identified among the evaluated channels, suggesting that the proposed system maintains adequate temporal consistency and can therefore be considered suitable for simultaneous multichannel sEMG acquisition applications.

\begin{table}
 \caption{Time per channel and differences between channels per event (in milliseconds, ms)}
  \centering
  \begin{tabular}{llllll}
    \toprule
    \cmidrule(r){1-6}
    Event & Channel 2 & Channel 4 & Channel 8 & $\Delta V$ (2 channel--4 channel) & $\Delta V$ (4 channel--8 channel) \\
    \midrule
    1 &	5318 &	5318 &	5318 &	0 &	0 \\
    2 &	10291 &	10291 &	10291 &	0 &	0 \\
    3 &	15809 &	15809 &	15809 &	0 &	0\\
    4 &	20700 &	20691 &	20691 &	9 &	0 \\
    5 &	25500 &	25491 &	25491 &	9 &	0\\
    6 &	31164 &	31164 &	31164 &	0 &	0\\
    7 &	36609 &	36609 &	36609 &	0 &	0\\
    8 &	44818 &	44818 &	44818 &	0 &	0\\
    9 & 52809 &	52800 &	52800 &	9 &	0\\
    10 &	54164 &	54164 &	54164 &	0 &	0\\
    11 &	55864 &	55864 &	55864 &	0 &	0\\
    \bottomrule
  \end{tabular}
  \label{tab:table5}
\end{table}

\begin{figure}[h]
  \centering
  \includegraphics[width=1\textwidth]{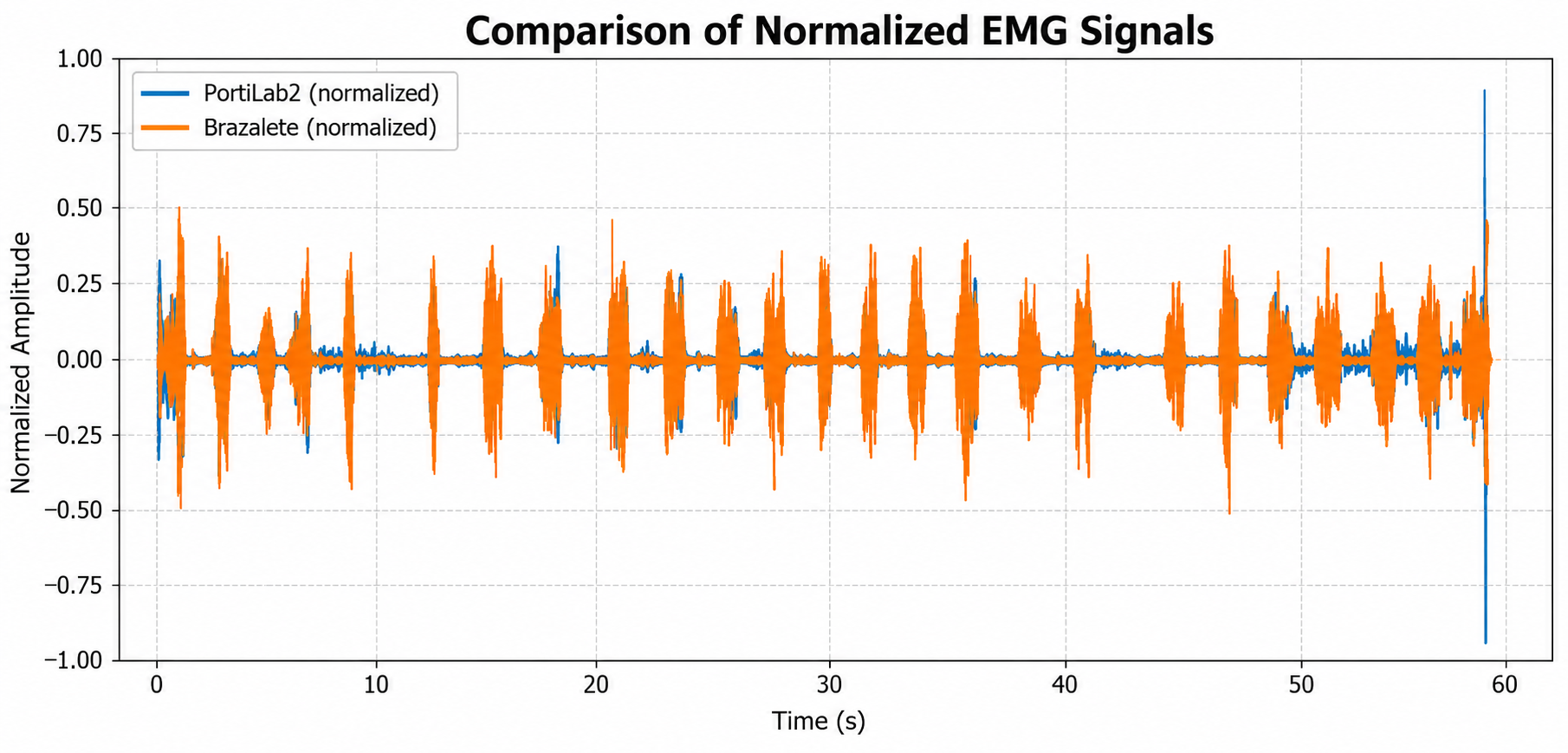}
  \caption{Visual comparison between PortiLab2 and Armband sEMG normalized signals.}
  \label{fig:fig6}
\end{figure}

Figure \ref{fig:fig6} illustrates the superposition of the normalized EMG signals acquired using the proposed prototype and the commercial reference system after the filtering and normalization stages. Overall, both signals exhibit a highly similar temporal behavior, particularly during the muscle activation intervals, where the amplitude variations and activation patterns closely coincide throughout most of the recording. This agreement suggests that the proposed system is capable of preserving the main characteristics of the EMG signal with behavior comparable to that of the reference device. In particular, slight temporal and amplitude variations become more evident in low-activity regions and near abrupt transitions, although these differences do not substantially alter the overall morphology of the recorded signal.

Complementarily, Table \ref{tab:table6} summarizes the quantitative results obtained for each extracted feature, providing a numerical comparison between both acquisition systems in terms of signal similarity and stability.

\begin{table}[h]
 \caption{Window-by-window comparison of EMG features (RMS, MAV, IEMG, VAR, WL) between the prototype and PortiLab2 using MAPE and Pearson’s correlation coefficient.}
  \centering
  \begin{tabular}{lll}
    \toprule
    \cmidrule(r){1-3}
    Metric & 1 – MAPE (\%) & Pearson correlation\\
    \midrule
    RMS (Root Mean Square) &	53.07 & 0.86\\
    MAV (Mean Absolute Value) &	54.45  & 0.91\\
    IEMG (Integrated Electromyography) &	54.46 & 0.91\\
    VAR (Variance) &	-7.09 & 	0.60\\
    WL (Waveform Length) &	55.77 & 	0.91\\
    \bottomrule
  \end{tabular}
  \label{tab:table6}
\end{table}

In general, the extracted electromyographic features demonstrated satisfactory agreement between the proposed prototype and the commercial reference system. Specifically, the root mean square (RMS), mean absolute value (MAV), integrated EMG (IEMG), and waveform length (WL) metrics achieved values of 1--mean absolute percentage error (1--MAPE) greater than 50\%, together with Pearson correlation coefficients above 0.85, indicating strong similarity in signal morphology, temporal activation patterns, and relative amplitude behavior. These results are consistent with the visual comparison presented in Figure \ref{fig:fig6}, where both normalized signals exhibit comparable waveform patterns throughout most of the recording.

\begin{figure}[h]
  \centering
  \includegraphics[width=1\textwidth]{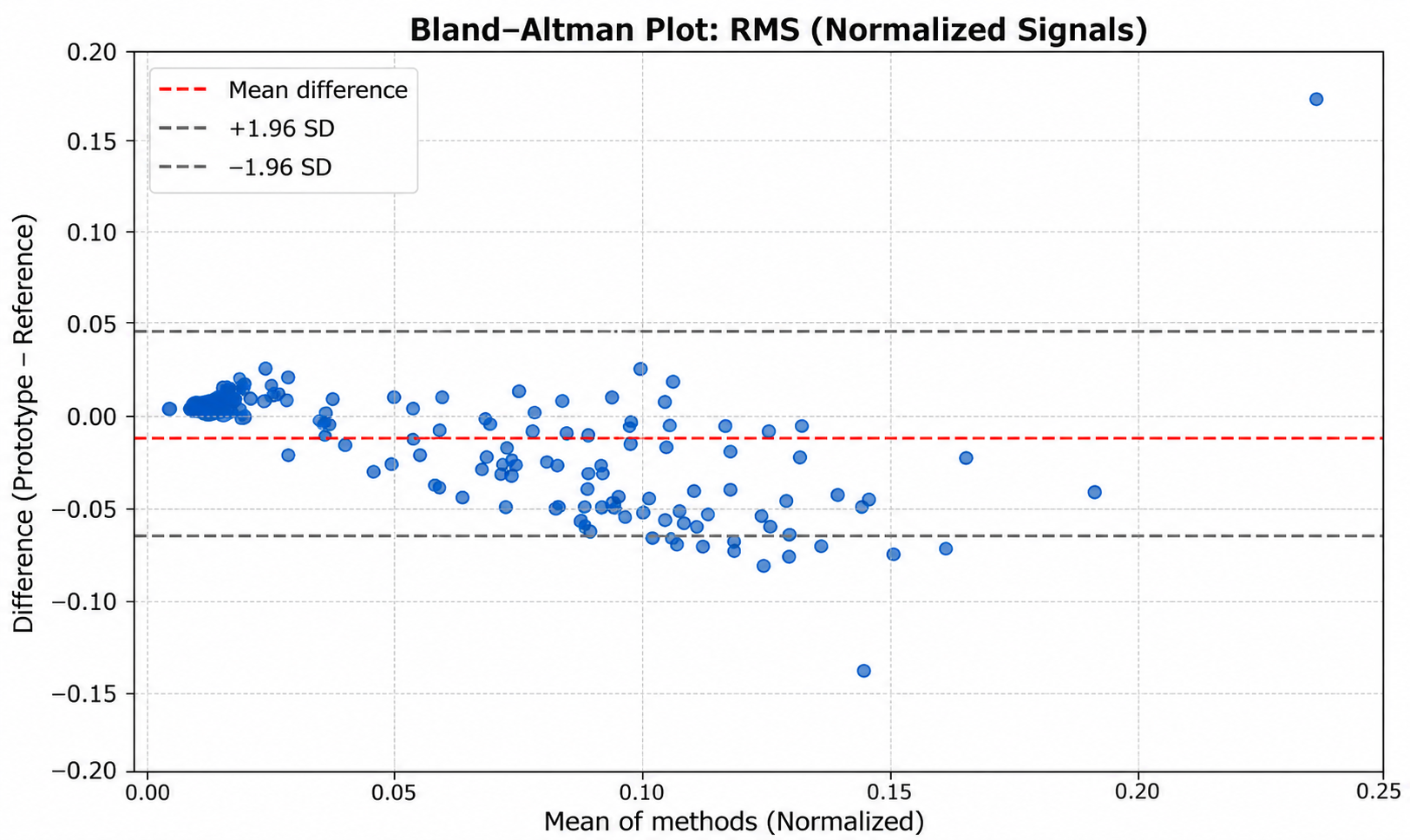}
  \caption{Bland–Altman plot applied to the comparison of the RMS metric.}
  \label{fig:fig7}
\end{figure}

In contrast, the variance (VAR) metric showed lower agreement, with reduced correlation and negative 1--MAPE values, indicating greater discrepancies between systems. This behavior is likely related to the sensitivity of variance to residual noise, baseline fluctuations, and small differences in amplification gain or filtering response. Since variance emphasizes signal dispersion, minor amplitude deviations may produce substantial changes in the calculated values. To complement the feature-based analysis, a Bland--Altman plot was used to evaluate agreement in the root mean square (RMS) metric between both signals, as shown in Figure \ref{fig:fig7}. The differences were distributed symmetrically around the mean difference, with most observations remaining within the $\pm$1.96 standard deviation limits and no evidence of systematic bias, supporting the overall agreement between the prototype and the commercial reference system.

\subsection{Communication}
For wired communication (UART), the device was connected to a computer via USB, and real-time data transmission was monitored. The acquired signals were correctly received and visualized without corrupted characters or structural inconsistencies. Additionally, during a continuous transmission period of 60 seconds, the total number of received samples matched the expected value based on the sampling frequency, indicating that no significant data loss occurred. The signal maintained temporal continuity, and no interruptions or gaps were observed.
 
Due to limitations in the Bluetooth Low Energy (BLE) module of the prototype, wireless validation was performed using an ESP32-based system implementing Bluetooth Classic as a reference. Under these conditions, data transmission remained stable, and the received signals preserved their structure and correspondence with the generated input. To evaluate robustness against interference, multiple Bluetooth devices (e.g., headphones and wireless peripherals) were activated simultaneously during transmission. Over a one-minute continuous test, no disconnections, delays, or data corruption were observed, suggesting adequate tolerance to electromagnetic interference in typical operating environments.
 
Furthermore, data integrity was assessed by verifying that the transmitted signals maintained their expected format and reflected variations consistent with muscle activity. The results confirm that the system preserves data integrity during transmission. Overall, the communication system demonstrated stable, loss-free performance in wired mode and robust behavior under wireless conditions when using a validated reference implementation.

\subsection{Comfort test}

During the resting condition, which consisted of 15 minutes of continuous use, participants did not report discomfort, localized pressure points, or skin irritation associated with contact between the device and the forearm. After device removal, only slight visible marks were observed on the skin, suggesting adequate pressure distribution and appropriate passive adaptation of the wearable structure to the user’s anatomy.

Similarly, during the active-use condition involving repetitive forearm movements such as flexion, extension, and rotation, the device maintained stable positioning without significantly restricting the natural range of motion. Participants did not report relevant discomfort related to friction, excessive compression, or displacement during movement. Nevertheless, among users with thinner forearms, a slight need for band readjustment was occasionally observed, suggesting that the current design could benefit from offering multiple band sizes or greater elastic stretch to improve fit across users.

\subsection{Mechanical compression assessment}

\begin{figure}[h]
  \centering
  \includegraphics[width=0.7\textwidth]{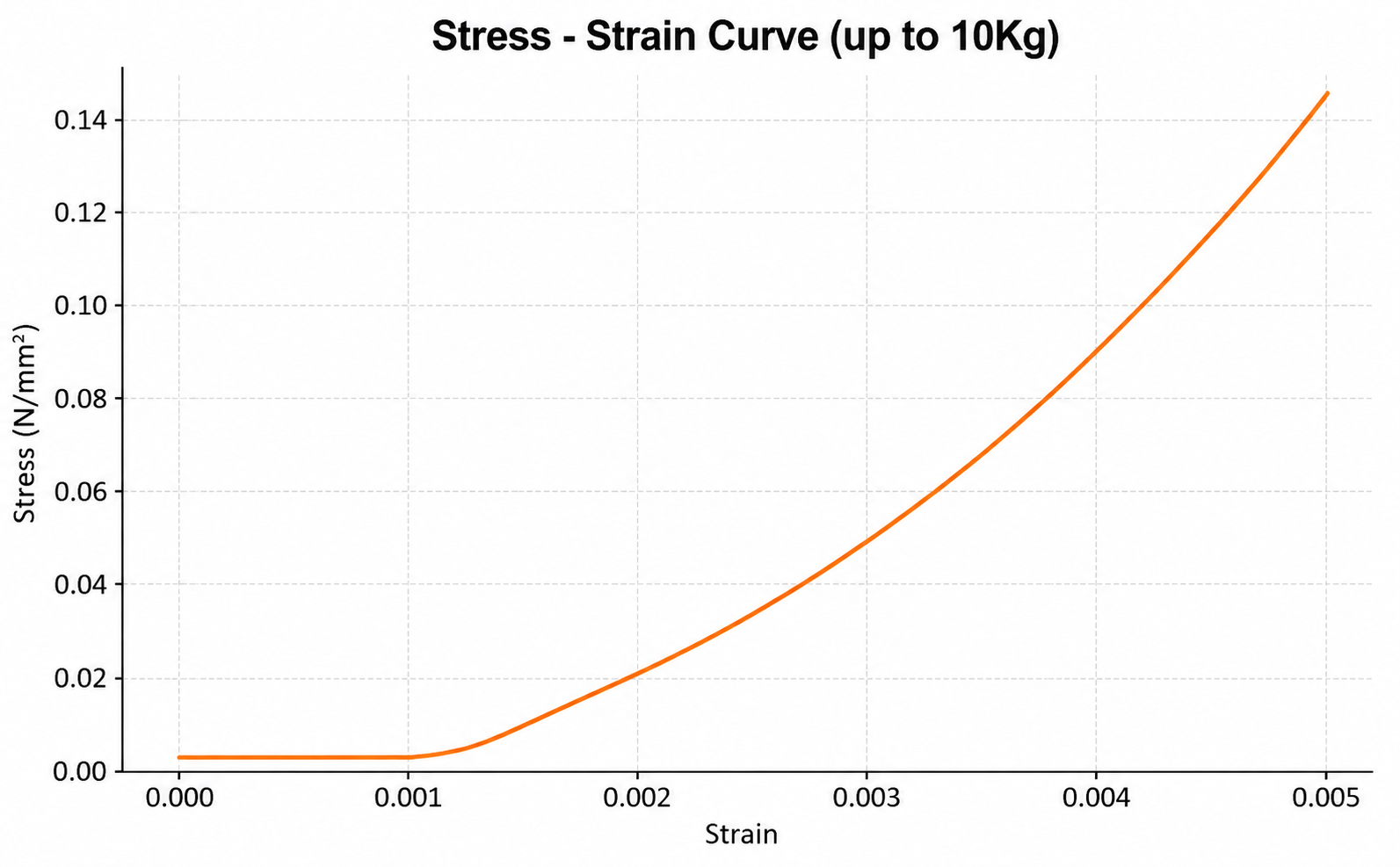}
  \caption{Stress–strain curve of the EMG sensor housing under a load of up to 10 kg.}
  \label{fig:fig8}
\end{figure}

The applied compressive load was progressively increased up to 98~N, equivalent to approximately 10~kg. Using the recorded force--displacement measurements, the corresponding stress--strain curve was constructed to characterize the mechanical response of the enclosure under compression. As illustrated in Figure \ref{fig:fig8}, the maximum stress reached during the test was approximately 0.15~MPa.

Throughout the entire loading range, the material exhibited predominantly linear elastic behavior, with no evidence of plastic deformation, structural instability, cracking, or mechanical failure after load removal. Furthermore, the maximum stress obtained during the experiment remained substantially lower than the typical yield strength reported for PETG, which ranges between 40 and 50~MPa. This considerable difference indicates the presence of a broad mechanical safety margin for the proposed enclosure design. 

\section{Discussion}
The application of the proposed validation protocol enabled a comprehensive assessment of electrical safety, signal quality, mechanical robustness, communication reliability, and usability of the developed sEMG armband. Beyond individual test results, the protocol provided a structured framework to interpret system performance holistically and to guide iterative design improvements.

From an electrical safety perspective, the measured leakage currents were slightly above the limits established by IEC 60601-1. Although the deviation was small and near the acceptable threshold, this result highlights the need for further refinement of the analog front-end and power supply design. In contrast, insulation tests confirmed adequate physical separation between the user and active circuitry, reducing the risk of direct electrical exposure. These findings illustrate how the protocol enables the identification of specific compliance gaps while confirming partial adherence to safety requirements.

The evaluation during operation demonstrated stable baseline behavior, with low variability and minimal internal noise when the system was tested without biological input. This indicates that the acquisition chain does not introduce significant interference under controlled conditions. However, discrepancies between simulated and experimental frequency responses were observed, particularly in the preamplification and filtering stages. These differences can be attributed to non-ideal characteristics of electronic components, parasitic effects, and implementation constraints, reinforcing the importance of experimental validation even when simulations are available.

Signal comparison with a commercial reference device showed a high correlation for amplitude-based features such as RMS, MAV, and waveform length, supporting the prototype's ability to capture meaningful muscular activity patterns.  Although a 1-MAPE of 90\%, typical of rigorous clinical validations, was not achieved. The obtained values are adequate for research-oriented prototype development. Some metrics, such as variance, exhibited lower agreement; these differences are consistent with the sensitivity of such features to noise and gain variations. Additionally, latency and interference tests confirmed that the system can operate effectively as a multichannel acquisition device, despite the sequential sampling architecture. These results suggest that the prototype is suitable for research-oriented applications, particularly in scenarios where relative signal behavior is more relevant than absolute precision.

The communication tests demonstrated robust performance in wired transmission, with no observable data loss, corruption, or temporal discontinuities during continuous acquisition. While limitations were identified in the BLE module, the use of an alternative Bluetooth implementation confirmed that the communication architecture is functionally sound. Furthermore, the system showed resilience to interference from nearby wireless devices, supporting its applicability in real-world environments. These findings highlight both the strengths of the current implementation and the need for hardware-level improvements in wireless communication.

From a mechanical standpoint, the enclosure exhibited purely elastic behavior under loads exceeding those expected in normal use, with stresses significantly below the material’s yield strength. The comfort assessment revealed that the device can be worn without significant discomfort during both static and dynamic conditions. Nevertheless, the need for improved adaptability to different forearm sizes was identified, indicating an opportunity for ergonomic refinement through adjustable or size-specific bands.

A key contribution of this work is the demonstration that a streamlined and accessible validation protocol can effectively support both device characterization and iterative development. By combining safety measurements, signal analysis, mechanical testing, communication validation, and usability assessment, the protocol enables researchers to identify concrete design improvements while simultaneously quantifying the reliability of the acquired data.

This dual role is particularly relevant in the context of sEMG database construction, where the quality of the dataset is directly dependent on the performance of the acquisition system. The results obtained—such as high correlation with a reference device, stable signal acquisition, and loss-free data transmission—provide confidence in the suitability of the prototype for generating research datasets. At the same time, identified limitations, including deviations in frequency response and marginal non-compliance in leakage current, establish clear boundaries for the interpretation of the data.

Overall, the proposed validation framework bridges the gap between early-stage prototyping and reliable experimental deployment. It enables the development of low-cost wearable systems with a clear understanding of their performance, limitations, and potential impact in human–machine interaction research. Future work will focus on improving electrical safety margins, optimizing wireless communication, and enhancing ergonomic adaptability, as well as extending the validation to long-term usage scenarios and larger user populations.

\paragraph{Acknowledgment} Research supported by Universidad de Antioquia, Colombia [CODI- PRG2024-77051, 2024].

\bibliographystyle{unsrt}  
\bibliography{references.bib}


\end{document}